\documentclass[runningheads]{svmult}
\usepackage{makeidx}   
\usepackage{epsfig}
\usepackage{graphicx}  
\usepackage{subeqnar}  
\usepackage{multicol}  
\usepackage{cropmark} 
\usepackage{physprbb}  
\newcommand{\simlt}
{\mbox{\raisebox{-0.5ex}{$\textstyle \; \sim$}
\raisebox{ 0.8ex}{$\textstyle  \!\!\!\!\!\!\! <$  }}}
\begin{document}
\title*{Dark Matter at the Center
and in the Halo of the Galaxy}
\toctitle{Dark Matter at the Center and
\protect\newline in the Halo of the Galaxy}
%
%
%
\titlerunning{Dark Matter at the Center}
%
\authorrunning{Neven Bili\'c et al.}
%
%
\author{Neven\,Bili\'c\inst{1,2} \and
Gary B.\,Tupper\inst{1} \and Raoul D.\,Viollier\inst{1}}
\institute{
Institute of Theoretical Physics and Astrophysics,
 Department of Physics, \\ University of Cape Town,
 Private Bag, Rondebosch 7701, South Africa
\and
Rudjer Bo\v skovi\'c Institute,
P.O.\ Box 180, 10002 Zagreb, Croatia}

\maketitle              

\begin{abstract}
All presently known stellar-dynamical constraints on the
size and mass of the supermassive compact dark object
at the Galactic center are consistent with a ball of
self-gravitating, nearly non-interacting, degenerate
fermions with mass between 76 and 491 keV/$c^{2}$,
for a degeneracy factor $g$ = 2.
Similar to the masses of neutron stars and
stellar-mass black holes, which are separated by an
Oppenheimer-Volkoff (OV) limit between 1.4 to 3
$M_{\odot}$, the masses of the supermassive fermion
balls and black holes are separated by an OV limit of
1.1 $\times$ 10$^{8} M_{\odot}$, for a fermion mass
of 76 keV/$c^{2}$ and $g$ = 2.

Sterile neutrinos of 76 keV/$c^{2}$ mass, which are
mixed with at least one of the active neutrinos with
a mixing angle $\theta \sim 10^{-7}$, are produced in
about the right amount in the early Universe by
incoherent resonant and non-resonant scattering of
active neutrinos having an asymmetry of $L \sim 10^{-2}$.
The former process yields sterile neutrinos with a
quasi-degenerate spectrum while the latter leads to
a thermal spectrum. The mixing necessarily implies the
radiative decay of the sterile neutrino into an active
neutrino in about 10$^{19}$ years which makes these
particles observable.

As the production mechanism of the sterile neutrino
is consistent with the constraints from large scale
structure formation, cosmic microwave background,
big bang nucleosynthesis, core collapse supernovae,
and diffuse X-ray background, it could be the dark
matter particle of the Universe. At the same time,
the quasi-degenerate components of this dark matter,
may be responsible for the formation of the
supermassive degenerate fermion balls and black
holes at the galactic centers via gravitational
cooling.
\end{abstract}
\section{Introduction}
In a recent paper Sch\"{o}del et al. reported
a new set of
data \cite{schod1}
including the corrected old
measurements \cite{eck2}
on the projected positions of the star S2(S0-2)
that was observed during the last decade with the
ESO telescopes in La Silla (Chile). The combined
data suggest that S2(S0-2)is moving on a
Keplerian orbit with a period of 15.2 yr around
the enigmatic strong radiosource Sgr A$^{*}$ that
is widely believed to be a black hole with a mass
of about 2.6 $\times$ 10$^{6} M_{\odot}$
\cite{eck2,ghez3}.
The salient feature of the new adaptive optics
data is that, between April and May 2002,
S2(S0-2) apparently sped past the point of
closest approach with a velocity $v$ $\sim$
6000 km/s at a distance of about 17 light-hours
\cite{schod1} or 123 AU from Sgr A$^{*}$.

Another star, S0-16 (S14), which was observed
during the last few years by Ghez et al.
\cite{ghez4}
with the Keck telescope in Hawaii, made
recently a spectacular U-turn, crossing the
point of closest approach at an even smaller
distance of 8.32 light-hours or 60 AU from
Sgr A$^{*}$ with a velocity $v$ $\sim$
9000 km/s. Ghez et al. [4] thus conclude
that the gravitational potential around
Sgr A$^{*}$ has approximately $r^{-1}$ form,
for radii larger than 60 AU, corresponding to
1169 Schwarzschild radii of 26 light-seconds
or 0.051 AU for a 2.6
$\times$ 10$^{6} M_{\odot}$ black hole.
Although the baryonic alternatives are
presumably ruled out, this still leaves
some room for the interpretation of the
supermassive compact dark object at the
Galactic center in terms of a finite-size
non-baryonic dark matter object rather
than a black hole.
In fact, the supermassive black hole
paradigm may eventually only be proven
or ruled out by comparing it with
credible alternatives in terms of
finite-size non-baryonic objects
\cite{mun9}.

The purpose of this paper is to explore,
using the example of a sterile neutrino
as the dark matter particle candidate,
the implications of the recent observations
for the degenerate fermion ball scenario of
the supermassive compact dark objects which
was developed during the last decade
\cite{mun9,viol5,viol6,bil7,bil8,mun10,bil11}.

\section{Stellar-dynamical constraints for fermion balls}

In a self-gravitating ball of degenerate
fermionic matter, the gravitational
pressure is balanced by the degeneracy
pressure of the fermions due to the
Pauli exclusion principle.
Nonrelativistically, this scenario is
described by the Lane-Emden equation
with polytropic index $p = 3/2$. Thus
the radius $R$ and mass $M$ of a ball
of self-gravitating, nearly non-interacting
degenerate fermions scale as \cite{viol6}
\begin{eqnarray}
\begin{array}{lcl}
R & = & \displaystyle{\left[ \frac{91.869 \;
\hbar^{6}}{m^{8} G^{3}} \; \left( \frac{2}{g}
\right)^{2} \; \frac{1}{M} \right]^{1/3}}\\[.5cm]
  & = & \displaystyle{3610.66 \; \mbox{ld}
  \left( \frac{15 \; \mbox{keV}}{mc^{2}}
  \right)^{8/3} \; \left( \frac{2}{g} \right)^{2/3} \;
  \left( \frac{M_{\odot}}{M} \right)^{1/3}} \; \; .
\end{array}
\end{eqnarray}
Here 1.19129 ld = 1 mpc = 206.265 AU, and $m$
is the fermion mass. The degeneracy factor
$g$ = 2 describes either spin 1/2
fermions (without antifermions) or spin
1/2 Majorana fermions. ($\equiv$
antifermions). For Dirac fermions and
antifermions, or spin 3/2 fermions
(without antifermions), we have $g$ = 4.
Using the canonical value
$M = 2.6 \times 10^{6} M_{\odot}$
and $R$ $\leq$ 60 AU for the supermassive
compact dark object at the Galactic center,
we obtain a minimal fermion mass of
$m_{\rm min}$ = 76.0 keV/$c^{2}$ for
$g$ = 2, or $m_{\rm min}$ = 63.9 keV/$c^{2}$
for $g$ = 4.

The maximal mass for a degenerate
fermion ball, calculated in a general
relativistic framework based on the
Tolman-Oppenheimer-Volkoff equations,
is the Oppenheimer-Volkoff (OV) limit
\cite{bil7}
\begin{equation}
M_{\rm OV} = 0.38322 \;
\frac{M_{\rm Pl}^{3}}{m^{2}} \,
\left( \frac{2}{g} \right)^{1/2}
=  2.7821 \times 10^{9} M_{\odot} \,
\left( \frac{15 \; \mbox{keV}}{mc^{2}}
\right)^{2} \, \left( \frac{2}{g}
\right)^{1/2}  ,
\end{equation}
where
$M_{\rm Pl} = (\hbar c/G)^{1/2} = 1.2210
\times 10^{19}$ GeV is the Planck mass.
Thus, for $m_{\rm min}$ = 76.0 keV/$c^{2}$
and $g$ = 2, or $m_{\rm min}$ = 63.9 keV/$c^{2}$
and $g$ = 4, we obtain
\begin{equation}
M_{\rm OV}^{\rm max} = 1.083 \times 10^{8} M_{\odot} \; \; .
\end{equation}
In this scenario all supermassive compact
dark objects with mass $M > M_{\rm OV}^{\rm max}$
must be black holes, while those with
$M \leq M_{\rm OV}^{\rm max}$ are fermion balls.

Choosing as the OV limit the canonical mass
of the compact dark object at the center of
the Galaxy,
$M_{\rm OV}^{\rm min}$ = 2.6 $\times$ 10$^{6}$ $M_{\odot}$,
yields a maximal fermion mass of
$m_{\rm max}$ = 491 keV/$c^{2}$
for $g$ = 2, or $m_{\rm max}$ =
413 keV/$c^{2}$ for $g$ = 4.
In this ultrarelativistic limit, there is
little difference between the black hole
and degenerate fermion ball scenarios, as the
radius of the fermion ball is 4.45 compared
to 3 Schwarzschild radii for the radius of
the event horizon of a non-rotating black
hole of the same mass. In fact, varying the
fermion mass between $m_{\rm min}$ and $m_{\rm max}$,
one can smoothly interpolate between a fermion
ball of the largest acceptable size and a fermion
ball of the smallest possible size, at the limit
between fermion balls and black holes.

The masses of the supermassive compact dark objects
discovered so far at the centers of both active and
inactive galaxies are all in the range
\cite{korm12}
\begin{equation}
10^{6} M_{\odot} \; \simlt \; M \; \simlt \;
3 \times 10^{9} M_{\odot} \; \; .
\end{equation}
Thus, as $M_{\rm OV}^{\rm max}$ falls into
this range as well, we need both supermassive
fermion balls
($M \leq M_{\rm OV}^{\rm max}$)
and black holes ($M > M_{\rm OV}^{\rm max}$)
to describe the observed phenomenology.
At first sight, such a hybrid scenario does not
seem to be particularly attractive. However, it
is important to note that a similar scenario is
actually realized in Nature, with the co-existence
of neutron stars which have masses
$M \leq M_{\rm OV}^{n}$, and stellar-mass black
holes with mass $M > M_{\rm OV}^{n}$, as observed
in stellar binary systems in the Galaxy
\cite{bland13}.
Here the OV limit
$M_{\rm OV}^{n}$, which includes the nuclear
interaction of the neutrons, is  somewhat uncertain
due to the unknown equation of state.
But the consensus of the experts \cite{bland13}
is that it must be in the range
\begin{equation}
1.4 \; M_{\odot} \; \leq \; M_{\rm OV}^{n} \;
\simlt \; 3 \; M_{\odot} \; \; .
\end{equation}
None of the observed neutron stars have masses
larger than 1.4 $M_{\odot}$, while there are at
least nine candidates for stellar-mass black holes
larger than 3 $M_{\odot}$ \cite{bland13}.
It is thus conceivable that Nature allows for the
co-existence of supermassive fermion balls and
black holes as well. Of course, we would expect
characteristic differences in the properties of
supermassive fermion balls and black holes.
Similarly, pulsars and stellar-mass black holes
are quite different, as pulsars have a strong
magnetic field and a hard baryonic surface,
while black holes are surrounded by an immaterial
event horizon instead. However, one may also argue
that the astrophysical differences between
supermassive black holes and fermion balls close
to the OV limit are not so easy to detect because
both objects are of non-baryonic nature.

\section{Cosmological constraints for
sterile neutrino dark matter}
If the supermassive compact dark object at the
Galactic center is indeed a degenerate fermion
ball of mass
$M$ = 2.6 $\times$ 10$^{6} M_{\odot}$ and radius
$R$ $\leq$ 60 AU, the fermion mass must be in the
range
\begin{eqnarray}
\begin{array}{c}
76.0 \; \mbox{keV}/c^{2} \; \leq \; m \; \leq \;
491 \; \mbox{keV}/c^{2} \; \; \mbox{for}
\; \; g = 2\\[.2cm]
63.9 \; \mbox{keV}/c^{2} \; \leq \; m \;
\leq \; 413 \; \mbox{keV}/c^{2} \; \;
\mbox{for} \; \; g = 4 \; \; .
\end{array}
\end{eqnarray}
It would be most economical if this particle
could represent the dark matter particle of
the Universe, as well.
In fact, it has been recently shown \cite{bil8}
that an extended cloud of degenerate fermionic
matter will eventually undergo gravitational
collapse and form a degenerate supermassive
fermion ball in a few free-fall times, if the
collapsed mass is below the OV limit. During
the formation, the binding energy of the nascent
fermion ball is released in the form of high-energy
ejecta at every bounce of the degenerate fermionic
matter through a mechanism similar to gravitational
cooling that is taking place in the formation of
degenerate boson stars  \cite{bil8}.
If the mass of the collapsed object is above
the OV limit, the collapse inevitably results
in a supermassive black hole.

The conjectured fermion could be a sterile
neutrino $\nu_{s}$ which does not participate
in the weak interactions.
We will now assume that its mass and degeneracy
factor is
$m_{s}$ = 76.0 keV/$c^{2}$ and $g_{s}$ = 2,
corresponding
to the largest fermion ball that is consistent
with the stellar-dynamical constraints. In order
to make sure that this fermion is actually
produced in the early Universe it must be mixed
with at least one active neutrino, e.g., the
$\nu_{e}$. Indeed, for an electron neutrino
asymmetry
\begin{equation}
L_{\nu_{e}} =
\frac{n_{\nu_{e}} - n_{\overline{\nu}_{e}}}
{n_{\gamma}} \sim 10^{-2}
\end{equation}
and a mixing angle $\theta_{es} \sim 10^{-7}$
\cite{abza14},
incoherent resonant and non-resonant active
neutrino scattering in the early Universe
produces sterile neutrino matter amounting
to the required fraction
$\Omega_{m} h^{2} = \left( 0.135_{- 0.009}^{+0.008}
\right)$ \cite{wmap15} of the critical density
of the Universe today. Here $n_{\nu_{e}}$,
$n_{\bar{\nu}_{e}}$ and $n_{\gamma}$ are the
electron neutrino, electron antineutrino and
photon number densities, respectively. An electron
neutrino asymmetry of $L_{\nu_{e}} \sim 10^{-2}$
is compatible with the observational limits on
$^{4}$He abundance, radiation density of the
cosmic microwave background at decoupling, and
formation of the large scale structure
\cite{kang16,espo17} which constrain the
electron neutrino asymmetry to the range
\begin{equation}
- 4.1 \times 10^{-2} \; \leq \; L_{\nu_{e}}
\; \leq 0.79 \; \; .
\end{equation}

Within these limits, $L_{\nu_{e}}$ must be
currently regarded as a free parameter which
may be determined by future observations,
in a similar way as the baryon asymmetry
\begin{equation}
\eta = \frac{n_{B} - n_{\overline{B}}}{n_{\gamma}}
\end{equation}
has been determined by big bang nucleosynthesis
to $\eta$ = (2.6 - 6.2) $\times$ 10$^{-10}$
\cite{fie18} and by the cosmic microwave
background radiation for
$\eta = \left( 6.1_{-0.2}^{+0.3}
\right) \times 10^{-10}$
\cite{wmap15}. There is no reason to expect
that $L_{\nu_{e}}$ and $\eta$ should be
of the same order of magnitude.

At this stage it is interesting to note that
incoherent resonant scattering of active
neutrinos produces
quasi-degenerate sterile neutrino matter,
while incoherent non-resonant active neutrino
scattering yields sterile neutrino matter that
has approximately a thermal spectrum
\cite{abza14}. Quasi-degenerate sterile neutrino
matter may contribute towards the formation of
the supermassive compact dark objects at the
galactic centers, while thermal sterile
neutrino matter is mainly contributing to the
dark matter of the galactic halos.

\section{Observability of degenerate sterile
neutrino balls}

The mixing of the sterile neutrino with at
least one of the active neutrinos necessarily
causes the main decay mode of the $\nu_{s}$
into three active neutrinos \cite{fie18} with a
lifetime of
\begin{equation}
\tau \left( \nu_{s} \rightarrow 3 \nu \right)
= \frac{192 \; \pi^{3}}{G_{F}^{2} \; m_{s}^{5}
\; \sin^{2} \;
\theta_{es}} = \frac{\tau (\mu^{-} \rightarrow
e^{-} + \bar{\nu}_{e} + \nu_{\mu})}{\sin^{2}
\theta_{es}} \;
\left( \frac{m_{\mu}}{m_{s}} \right)^{5} \; \; ,
\end{equation}
which is presumably unobservable as the available
neutrino energy is too small. Here
$\tau (\mu^{-} \rightarrow e^{-} + \bar{\nu}_{e}
+ \nu_{\mu})$ and $m_{\mu}$ are the lifetime and
mass of the muon. However, there is a subdominant
radiative decay mode of the sterile into an active
neutrino and a photon with a branching ratio
\cite{pal20}
\begin{equation}
B (\nu_{s} \rightarrow \nu \gamma) =
\frac{\tau (\nu_{s} \rightarrow 3 \nu)}{\tau
(\nu_{s} \rightarrow \nu \gamma)}
= \frac{27 \; \alpha}{8 \pi} = 0.7840 \times 10^{-2} \; \; ,
\end{equation}
where $\alpha = e^{2}/\hbar c$ is the fine
structure constant. The lifetime of this potentially
observable decay mode is
thus
\begin{equation}
\tau \left( \nu_{s} \rightarrow \nu \gamma \right)
= \frac{8 \pi}{27 \; \alpha} \;
\frac{1}{\sin^{2} \theta_{es}} \;
\left( \frac{m_{\mu}}{m_{s}} \right)^{5} \;
\tau \; (\mu^{-} \rightarrow e^{-} +
\bar{\nu}_{e} + \nu_{\mu})
\end{equation}
yielding, for $\theta_{es} = 10^{-7}$ and
$m_{s}$ = 76.0 keV/$c^{2}$, a lifetime of
$\tau (\nu_{s} \rightarrow \nu \gamma) =
0.46 \times 10^{19}$ yr.

Although the X-ray luminosity due to the
radiative decay of diffuse sterile neutrino dark
matter in the Universe is presumably not observable,
because it is well below the X-ray background
radiation at this energy \cite{abza14}, it is
perhaps possible to detect such hard X-rays in
the case of sufficiently concentrated dark matter
objects.
In fact, this could be the smoking gun for both
the existence of the sterile neutrino and the
fermion balls. For instance, a ball of $M = 2.6
\times 10^{6} M_{\odot}$ consisting of degenerate
sterile neutrinos of mass
$m_{s}$ = 76.0 keV/$c^{2}$ \cite{mun10},
degeneracy factor $g_{s}$ = 2, and mixing angle
$\theta_{es}$ = 10$^{-7}$
would emit 38 keV photons with a luminosity
\begin{equation}
L_{X} = \frac{M c^{2}}{2 \tau (\nu_{s}
\rightarrow \nu \gamma)} = 1.6 \times 10^{34} \;
\mbox{erg/s} \; \; ,
\end{equation}
within a radius of 60 AU, 8.32 light hours or
7.6 $\times$ 10$^{-3}$ arcsec of Sgr A$^{*}$,
assumed to be at a distance of 8 kpc. The
current upper limit for X-ray emission from the
vicinity of Sgr A$^{*}$ is
$\nu L_{\nu} \sim 3 \times 10^{35}$ erg/s,
for an X-ray energy of $E_{X} \sim 60$ keV
\cite{maha21}, where $L_{\nu} = dL/d \nu$ is
the spectral luminosity. Thus the X-ray line at
38 keV could presumably only be detected if
either the angular or the energy resolution
or both, of the present X-ray detectors are
increased.

\end{document}